\documentclass[letterpaper, 10 pt, conference]{ieeeconf}  % Comment this line out if you need a4paper

\IEEEoverridecommandlockouts                              % This command is only needed if 
\usepackage{cite}
\usepackage{amsmath}
\usepackage{amssymb}
\usepackage{array}
\usepackage{subfiles}
\usepackage{algorithm}

\usepackage{algorithmic}
\usepackage{graphicx}
\usepackage{textcomp}
\usepackage{romannum}
\usepackage{subcaption}

\newtheorem{definition}{Definition}
\newtheorem{lemma}{Lemma}
\newtheorem{theorem}{Theorem}

\title{\LARGE \bf
	Potential Game-Based Non-Myopic Sensor Network Planning for Multi-Target Tracking
}

\author{Su-Jin~Lee, Soon-Seo~Park and Han-Lim Choi% <-this % stops a space
	\thanks{{\tt\small $\{$sjlee, sspark, hanlimc$\}$@lics.kaist.ac.kr}}%    
}

\begin{document}

	\maketitle
	\thispagestyle{empty}
	\pagestyle{empty}

\begin{abstract}
	%This paper presents a potential game approach for non-myopic planning of a mobile sensor network in order to gather information about the moving multiple targets as much as possible over a planning horizon. The constraints on the sensing platform motion and limitations in sensor capabilities cause This long-term planning is suffered from exponential increase in computation time, 
	%minimize overall tracking error at the last planning time step
	\iftrue
	This paper presents a potential game-based method for non-myopic planning of mobile sensor networks in the context of target tracking. The planning objective is to select the sequence of sensing points over more than one future time steps to maximize information about the target states. This multi-step lookahead scheme is studied to overcome getting trapped at local information maximum when there are gaps in sensing coverage due to constraints on the sensor platform mobility or limitations in sensing capabilities. However, the long-term planning becomes computationally intractable as the length of planing horizon increases. This work develops a game-theoretic approach to address the computational challenges. The main contributions of this paper are twofold: (a) to formulate a non-myopic planning problem for tracking multiple targets into a potential game, the size of which linearly increases as the number of planning steps (b) to design a learning algorithm exploiting the joint strategy fictitious play and dynamic programming, which overcomes the gaps in sensing coverage. Numerical examples of multi-target tracking demonstrate that the proposed method gives better estimation performance than myopic planning and is computationally tractable. 
	\fi
	\iffalse
	This paper presents a potential game-based method for non-myopic planning of sensor networks in the context of target tracking. The main contributions of this paper are twofold: (a) to formulate a non-myopic planning problem for tracking multiple targets into a potential game, the size of which linearly increases as the number of planning steps (b) to design a learning algorithm exploiting the joint strategy fictitious play and dynamic programming, which overcomes the gaps in sensing coverage. Numerical examples of multi-target tracking demonstrate that the proposed method gives better estimation performance than myopic planning and is computationally tractable. 
	\fi
	\iffalse
	algorithm for a multi-target tracking problem. %The planning objective is to select the sequence of sensing points over greater than one future time steps to maximize information about the target states. 
	This multi-step lookahead scheme is studied to overcome getting trapped at local information maximum when there are gaps in sensing coverage due to constraints on the sensor platform mobility or limitations in sensing capabilities. However, the long-term planning becomes computationally intractable as the number of time steps that the algorithm plans. This work develops a game-theoretic approach to address the computational burden by extending the potential game formulation proposed by the present authors. T
	\fi
\end{abstract}

\maketitle

\section{Introduction} 
\label{sec:introduction}

% Subject: Sensing points planning for mobile sensor network
% Topic: What is the efficient method to find out the sensing points for mobile sensor network? 
% Conclusion: Game-theoretic approach to non-myopic planning is the efficient method 

% Question1: What is the efficient method to find out the path for multiple robots that maximizes the information about the environment? 
% Question2: Why is the proposed method efficient? 

%\IEEEPARstart{M}{obile}
%The goal of sensor network planning is to determine the best sensing locations so that the information gain about the quantities of interest are maximized. Generally, each sensor has some limitations on its mobility, communication range and/or field of view about the sensing area. 

%When monitoring wide geographic areas for tasks, such as tracking targets,  weather forecast, and mapping, using a mobile sensor network, determining where to take measurements for a number of sensors is a fundamental problem\cite{Singh2009_IJCAI, Choi2011_IEEETCST, La2015_IEEETSMCS, Hoffmann2010_IEEETAC, Kreucher2004_IEEECDC}. 

% What we are going to do in this paper
 Sensor networks have been successfully used to acquire information about the quantities of interest spread over large areas, including applications such as, monitoring spatial phenomena, mapping, and tracking targets \cite{Singh2009_IJCAI, Choi2011_IEEETCST, Choi2011_Sensors, La2015_IEEETSMCS, Hoffmann2010_IEEETAC, Kreucher2004_IEEECDC}. The mobility of the sensor platforms certainly contribute to expanding the sensing coverage and mission areas, the associate constraints in mobility and sensor modalities necessitates resolution of another crucial decision making on where and when to sense with taking into account the constraints. The waypoints of the sensor platforms need to be chosen to maximize the information gain while satisfying all the associated resource constraints. 

% Introdue two approaches for solving the problem: myopic and myopic 
%	Myopic
%	Limitations in myopic
% 	Therefore, non-myopic   , Gostar2017_IEEETAES
Sensor planning schemes may be myopic or non-myopic in terms of time domain. When we consider a set of sensing locations for the next time step only, this approach is referred to as greedy or myopic. This short term management has given good performance in many tasks mentioned earlier (e.g., sensor placements for monitoring spatial phenomena \cite{Krause2008_JMLR}, mobile sensor targeting for weather forecast \cite{Choi2011_IEEETCST, Choi2011_Sensors}, exploration path generation for simultaneous localization and mapping \cite{Kretzschmar2012_IJRR,yang2018optimal}, sensor management for target tracking \cite{Hoffmann2010_IEEETAC, Choi2013_OptmLet, Bogdanovic2016_arXiv, Farmani2017_IEEETAES, jiang2018online}. However, there are some situations in which non-myopic strategies give poorer performance in the next step, but better  estimation accuracy at the end of the planning horizon. In \cite{Kreucher2004_IEEECDC}, \cite{Liu2003_ICASSP}, and \cite{Ding2012_ACC}, multi-step lookahead sensor planning was shown to significantly improve the performance for target tracking. Typically, when a sensor network contains "sensor holes", a greedy algorithm results in poor performance \cite{Liu2003_ICASSP}. The sensor holes are gaps in the sensing coverage where the sensor network cannot make observations of the targets. These invisible areas are caused by the limited field of view of sensors or by the obstacles existing in environments such as elevation difference of surveillance region or by physical constraints on the sensor platform motion. This is the case for tracking moving objects with such sensors of a restricted capability. Non-myopic planning can address this sensor hole issue by looking multiple steps ahead in the future to extend candidate sensing regions.

%We propose a game-theoretic approach to address non-myopic planning for multi-target tracking. 

% What effors has been done to solve the non-myopic planning
% 	Address our previous work on myopic planning using a potential game formulation
This paper considers non-myopic planning for multi-target tracking. % is addressed by applying a game-theoretic approach to find the optimal path efficiently. 
Previous works on sensor planning for target tracking can be found in \cite{Chhetri2006_JASP, Williams2007_IEEETSP, Singh2009_IJCAI, Oh2013_MED}, and resource management for multi-function radar \cite{Bogdanovic2016_arXiv, Charlish2012_ICIF}. Non-myopic sensor planning is difficult because the number of possible sensing sequences increases exponentially with the length of the planning horizon. Exponential explosion in computational time and memory usage follow for finding the optimal sensing sequence. Several planning algorithms have been proposed to reduce the computational costs. A simple but suboptimal strategy is a greedy method that determines sensing points in sequence, choosing the next location which provides the maximum information gain conditioned on the preceding sensing decisions \cite{Singh2009_IJCAI}. Approaches to find the optimal solutions can be found in \cite{Logothetis1999_ACC, Chhetri2006_JASP, Williams2007_IEEETSP}. The methods used approximate cost functions and reduced the computational burden by using pruning algorithms. Since the sensor planning has the uncertain nature of the underlying states, adaptive path planning problems have been formulated as a Partially Observable Markov Decision Process (POMDP) \cite{Castanon1997_IEEECDC}. \cite{Chong2009_POMDP} designed the guidance algorithms for controlling unmanned aerial vehicles by solving a POMDP approximately. However, these more complex algorithms still suffer from the computation cost, thus can deal with only the limited size of the problems. %TODOv: Add Greedy (Choid) POMDP, game theoretic method, dynamic programming(Williams)  - one approach / another approach structure ?

% What approaches are used to get the solution for non-myopic planning
% 	Potential game 
%	Mutual information 
% 	
The work addresses the computational challenge by formulating a non-myopic planning problem as a potential game. %The goal of the sensor planning problem considered here is to find the informative paths for mobile sensor networks over the multiple time steps. 
The method is extended from the potential game formulation that was proposed in the authors' earlier work \cite{Choi2013_ACC}, \cite{Choi2015_IEEETCST} for efficient selection of sensing points for the next time step only. To formulate an optimization problem into a potential game, we need to specify players, their respective actions, their local objective/utility functions, and the learning rules for the players \cite{Gopalakrishnan2011_ACM}. In the game framework of \cite{Choi2015_IEEETCST}, each sensor in a sensor network represents a player trying to maximize its utility function defined by the conditional mutual information.  This local utility design leads to a potential game, with a global objective being the mutual information between the target states and the measurement variables to be taken at the next time step. In addition to our earlier work, there are more works using game theory for distributed sensor network management. For example, track selection for multi-target tracking in a multifunction radar network has been made through an anti-coordination game \cite{Bogdanovic2016_arXiv}. A radar in a network tries to maximize the overall tracking accuracy criterion and a best response based dynamics is presented to find an equilibrium. In \cite{Shen2010_SPIE}, the sensor management assigns moving sensors to targets through a distributed sensor-based negotiation game. Each sensor makes a decision locally that maximizes its utility by negotiating with the neighboring sensors. In \cite{Panoui2016_IEEETAES}, clusters of multistatic radars in a network select the optimal waveforms maximizing the signal-to-disturbance ratio through a potential game. The authors proved the uniqueness of the Nash equilibrium using the discrete concavity property of the proposed game.  However, these game-theoretic approaches also have the same computational cost issue when considering long-term planning, thus these studies handled short-term management only.  % 우리는 기존의 potential game formulation을 두 개의 

To yield an efficient potential game method for informative non-myopic planning, we modify the game framework in our earlier work and propose a learning algorithm. In formulating a sensor network planning problem into a potential game, we take three steps: (\romannum{1}) specify the global objective function for multi-target tracking, (\romannum{2}) formulate the optimization problem as a potential game. (\romannum{3}) design a learning algorithm to find a Nash equilibrium of the potential game. First, mutual information is used as a global objective of a planning problem. The quantities of interest are set to the target states at the last time of the planning horizon to reduce the computational burden. Then, a potential game is formulated by defining a player as a sensor at each time step. A local utility is defined by using the marginal contribution rule \cite{Choi2015_IEEETCST}, \cite{Marden2013_OR}, it is shown to lead to a potential game with the mutual information. Lastly, we propose a learning method to find a solution of the designed game, which is extended from the joint strategy fictitious play \cite{Marden2009_IEEETAC}. Dynamic programming is used in the first run of the learning algorithm in order to fill the gaps in sensing coverage caused by the limited sensing capabilities and constraints on platform mobility. Numerical studies on UAV coordination for optimal ground target tracking are presented to demonstrate that the game-theoretic mechanisms provide computational efficiency in sensor planning over multiple time steps.  

% 풀어야할 문제는 combinatorial optimization 문제이다. 따라서 이를 해결하기 위해서는 두 가지 

%In summary, the main contributions of this paper are two-fold. First, this paper propose a potential game framework to 

% 그러나 넓은 지역을 해야하고, limitations on resource가 있기 때문에 최적화가 필요하다. 

% 센서 네트워크(UAV)가 요즘 많이 사용되어지고 있다. such as environmental monitoring, security surveillacne, battlefield awareness [\cite{Liu2003_ICASSP}]
% 
% 센서 네트워크 플래닝은 관심 변수에 대해서 정보를 얻으내는 문제이므로 최적화 문제이고, 따라서 두 가지 관점에서 많은 연구가 진행되었다. in myopic planning 
% 첫째는 여러가지 인수의 combination에 대한 목적함수의 value를 효율적으로 계산하기 위한 연구가 있었고, (Choi-Backward scheme)
% 두번째는 계산해야할 combination 자체의 수를 줄이기 위한 연구가 있었다. (Choi - greedy, potential, Krause). 
% 그 중에서도 센서의 limited field of view가 있는 경우에는 non-myopic을 하는데, 
% 
% Myopic planning is greedy in the temporal sense.
%  myopic planning은 일반적으로 좋은 결과를 주지만, sensor hole과 같이 센서가 limited field of view가 있는 경우에는 local optimum에 빠질 수 있다. [\cite{Liu2003_ICASSP}-fixed wireless sensor network에서의 limited sensing area for the next time step is similar to the limited field of view in mobile sensor network ] shows a toy problem where... 이것을 극복하기 위해서 non-myopic planning 이 많은 ��%�문에서 여러가지 상황에 대해서 제안되었다. 
% myopic planning은 시간에 대해서 greedy하기 때문에 sensor가 limited field view를 가지고 있는 경우에는 한정된 정보(no global knowledge)만을 사용한다. 

%TODO: Add the result for the myopic result and non-myopic simulation results that has similar performance.
%\IEEEpubidadjcol

\section{Problem Formulation}
%In this section, we define the multi-target tracking problem with a mobile sensor network and formulate a planning problem to find out the sensing locations to enhance the tracking performance. First, the models of components comprising the target tracking problem are defined, then a information-theoretic framework for planning is proposed using mutual information as an objective function to optimally locate the sensing agents. 
%The target tracking problem is to estimate the kinematic states of targets with a finite set of measurements. In general, the state estimation is performed via the Bayesian filtering \cite{Chen2003_Stat} which is a recursive algorithm consisting of two processes: predicts the prior distribution of the target states using the dynamic model of the target and updates the distribution with sensor measurement model and taken measurements. So, specific models of target dynamics and sensor measurement are presented for the sake of concreteness. However, note that the planning algorithms to be introduced can be readily applied to other problems in which sensors have a finite set of actions (i.e., sensing locations, sensor modes) to select. 
While the game-theoretic method of cooperative planning applies to many different applications with sensor networks, here we illustrate the sensor planning in the context of multi-target tracking. 

In this section, we define a multi-target tracking problem with a mobile sensor network and formulate a planning problem to find out the sensing locations to enhance the tracking performance. First, the models of components comprising the target tracking problem are defined, then a data fusion method for planning is described briefly to represent the conditional probability density function (pdf) of the target state conditioned on the measurements. 

The goal of a target tracking problem is to estimate the kinematic states of targets with a finite set of measurements. In general, the state estimation is performed via the Bayesian filtering \cite{Chen2003_Stat} which is a recursive algorithm consisting of two processes: predicts the prior distribution of the target states using the dynamic model of the target and updates the distribution with sensor measurement model and taken measurements. So, specific models of target dynamics and sensor measurement are presented for the sake of concreteness. However, note that the planning algorithms to be introduced can be readily applied to other problems in which sensors have a finite set of actions (i.e., sensing locations, sensor modes) to select. 

\begin{figure}[t]
	\centerline{\includegraphics[width=1\columnwidth]{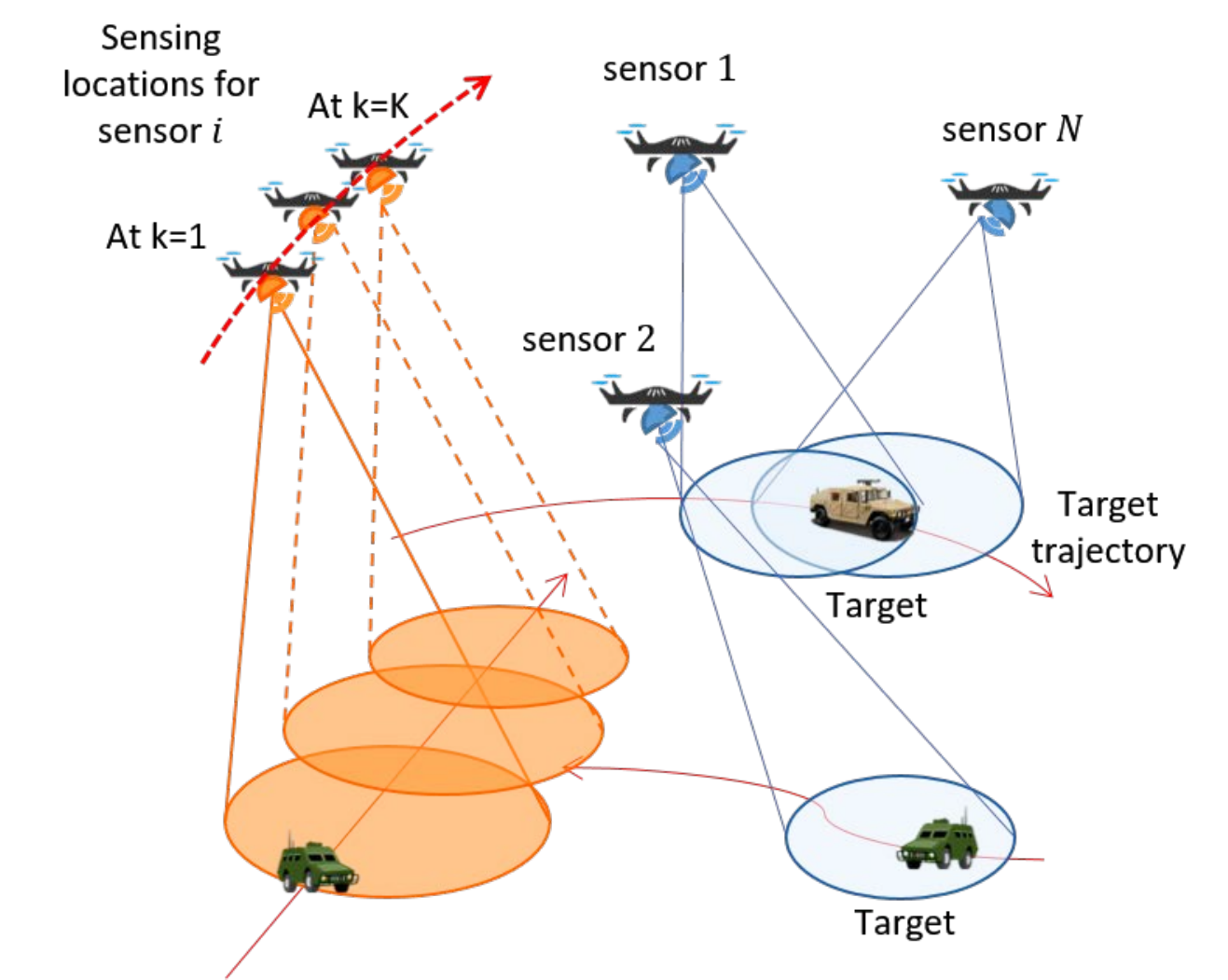}}
	\caption{A sensor network planning problem for multi-target tracking: A number of targets moving on the ground are tracked by a sensor network. Every UAV (unmanned aerial vehicle) is carrying a sensor which observes the targets at a prespecified height. } 
	\label{fig:sec2_prob_setting}
\end{figure}

\subsection{Target and Sensor Models} %Multi-target Tracking Problem: 
We consider a number of Unmanned Aerial Vehicles (UAVs) carrying sensors to track a group of targets on the ground as shown in Fig. \ref{fig:sec2_prob_setting}. To estimate the target states with measurements, a sensor network keeps information about the targets using a motion model and a prior distribution of the states. 

Let a set of targets moving on the ground to be $\mathcal{T}=\{1,2,\dots,M\}$. 
Here, we assume that the number of targets are known and constant as $M$.  %Searching is not covered in this paper 
The state of target $j\in \mathcal{T}$ at time $k$ is denoted by $\mathbf{x}_k^{(j)}=[x_k^{(j)}, y_k^{(j)}, \dot{x}_k^{(j)}, \dot{y}_k^{(j)}]^T$, where $x_k^{(j)}$ and $y_k^{(j)}$ represent the target position in two dimensional Cartesian coordinates, and $\dot{x}_k^{(j)}$ and $\dot{y}_k^{(j)}$ are the corresponding velocity components. We assume that each target follows continuous white noise acceleration models and moves independently from others, then it is possible to ignore the joint distribution and use one tracking filter for each target \cite{Skoglar2007_Book}.
\begin{equation}
\label{eq:model_targetdyn}
\mathbf{x}_{k+1}^{(j)} = F_k\mathbf{x}_k^{(j)} + w_k^{(j)}
\end{equation}
where $w_k^{(j)}\sim\mathcal{N}(0, Q_k^{(j)})$ is a white Gaussian process noise, independent of the other targets and measurements. $F_k$ and $Q_k$ are system transition and process noise covariance matrices, respectively. For the simulations in this paper, we use a continuous white noise acceleration model. 
\begin{equation}
F_k=\begin{bmatrix}
1 & 0 & \Delta t & 0  \\
0 & 1 &         0 & \Delta t \\
0 & 0 & 1 & 0 \\
0 & 0 & 0 & 1
\end{bmatrix}; ~~
Q_k = \begin{bmatrix}
\frac{\Delta t ^3} {3} & 0 & \frac{\Delta t^2} {2} & 0  \\
0 & \frac{\Delta t ^3} {3}&         0 &  \frac{\Delta t^2} {2} \\
\frac{\Delta t^2} {2} & 0 & \Delta t & 0 \\
0 & \frac{\Delta t^2} {2} & 0 & \Delta t
\end{bmatrix} q
\label{eq:model_targetdyn_cv}
\end{equation}
where $\Delta T$ is the time between two successive measurements at $k$th step to $(k+1)$th step and $q$ is the process noise intensity, representing the strength of the deviations from predicted motion by the dynamic model \cite{Charlish2011_Thesis}. When $q$ is small, this model represents a nearly constant velocity. Note however, the actual maneuver of the targets can be different from the dynamics predicted by the sensor network. In the simulations, we will show the effect of incorrectly fitted models on performance of planning algorithms. 

\begin{figure}[t]
	\centering
	\hspace*{-0.5cm}%
	\includegraphics[width=.53\columnwidth]{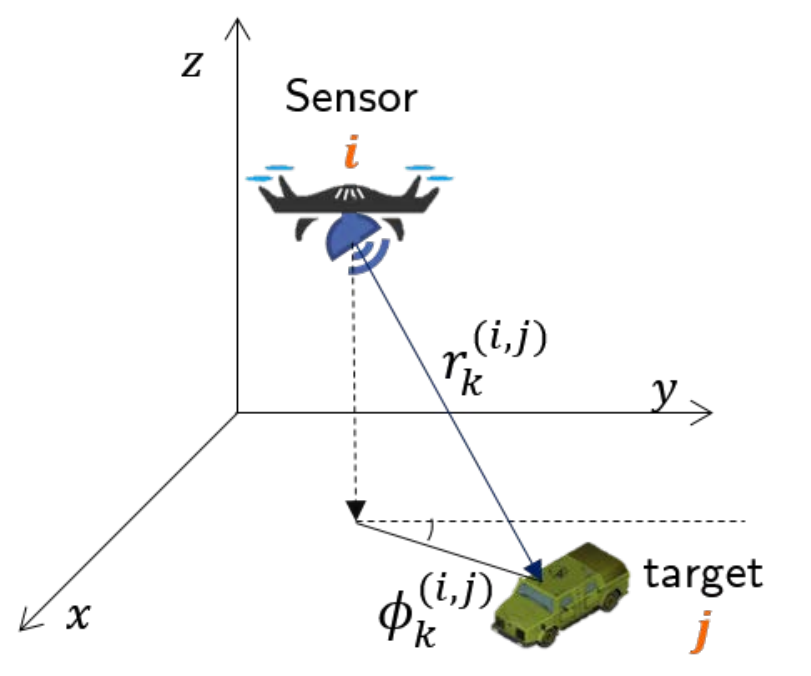}
	\hspace*{-0.3cm}%
	\includegraphics[width=.53\columnwidth]{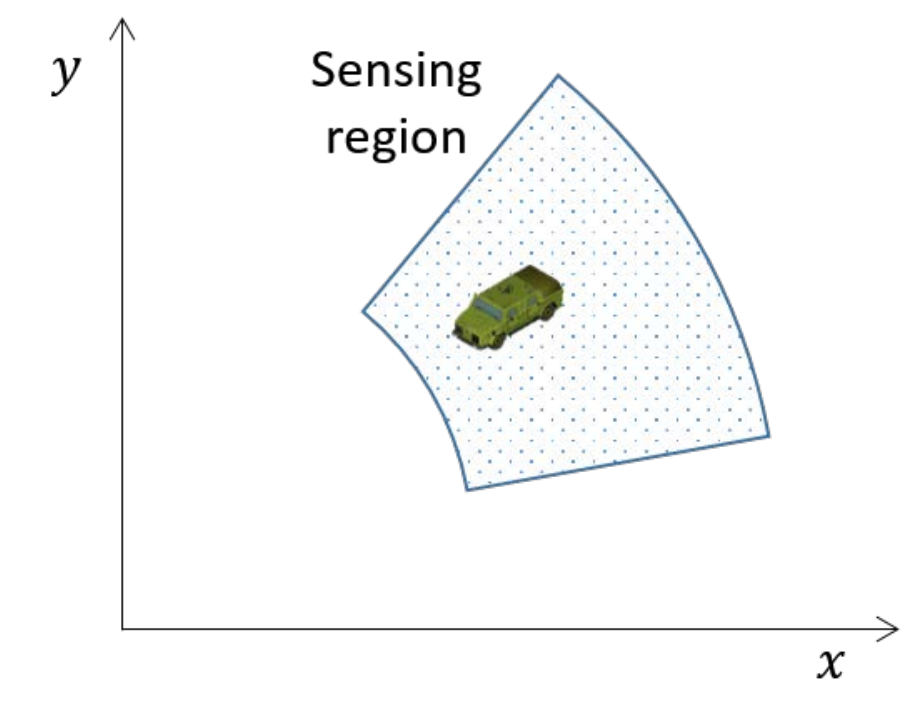}
	\caption{Radar-like measurement model for target tracking. (left) A sensor is mounted on an UAV and takes measurements consisting of range and azimuth to the target. (right) Sensors have limited sensing region, in which a target can be detected and the relative position of the target can be measured.}
	\label{fig:sec2_radarmodel}
\end{figure}

A sensor network is represented by a set of vehicles $\mathcal{N}=\{1,2,\dots,N\}$, each vehicle is assumed to be equipped with one sensor for simplicity\footnote{In this paper, we use a sensing agent, a sensor, and a sensing node interchangeably. For simplicity, we assume that each vehicle have one sensor on it. In general, some platforms can carry more than one sensor. In that case, the sensing locations for the sensors equipped on the same vehicle should match. Then the decision variables are for the vehicles not for the sensors.}.  For each sensor $i\in\mathcal{N}$, the measurement taken at time $k$ is denoted in a general nonlinear form as 
\begin{equation}
z_k^{(i)} = h_k^{(i)}(\mathbf{x}_k)+v_k^{(i)}
\end{equation}
where $v_k^{(i)} \sim \mathcal{N}(0,R_k^{i})$ is a white Gaussian noise process, independent of the other measurement noises and of process noise $w_k^{(j)},~\forall k,~\forall j\in\mathcal{T}$. $\mathbf{x}_k$ is the set of states of targets and a sensor network. More specifically, in a target tracking problem, sensors usually measure the kinematic information about the target relative to the sensor, itself. Thus the state of the $i$-th sensor in $\mathbf{x}_k$  includes position, orientation, and velocity of the agent. Denoting the pose of $i$-th sensing agent at time $k$ as $\mathbf{x}_k^{(i)_s}$, the measurement model can be rewritten to express the relative kinematic state of the target as 
\begin{equation}
z_k^{(i)} =h(\mathbf{x}_k^{(j)},~\mathbf{x}_k^{(i)_s})+v_k^{(i)}
\end{equation}
For the simulations in this paper, we set the measurement model to a \textit{radar-like measurement}, consisting of range and azimuth to a target.  At each time $k$, sensor $i$ obtains the positional measurement $\mathbf{z}_k^{(i,j)}=[r_k^{(i,j)},~ \phi_k^{(i,j)}]^T$ for one of the targets, $j\in\mathcal{T}$: 
\begin{equation}
\begin{bmatrix}
r_k^{(i,j)} \\
\phi_k^{(i,j)}
\end{bmatrix}  = \begin{bmatrix}
\sqrt{(x_j-x_i)^2 + (y_j-y_i)^2 + z_i^2} \\
\tan^{-1}((y_j-y_i)/(x_j-x_i))
\end{bmatrix}.
\end{equation}
which is illustrated in Fig \ref{fig:sec2_radarmodel}. For notational simplicity, the discrete time index $k$  can be dropped later. The superscripts $i$ and $j$ represent the index of target and sensor, respectively, in the rest of this paper. As shown in Fig. \ref{fig:sec2_radarmodel}, sensors have directivity, thus only the target inside the sensing region can be detected and the sensor can obtain the information about the location of the target within that area. 

In radars it is shown that the signal to noise ratio (SNR) dependent error is a major factor of the measurement error \cite{SHKim2013_Thesis}. An azimuth angle error can be modeled as a white Gaussian noise with zero mean and standard deviation given by 
\begin{equation}
\sigma_\phi \approx \frac{\theta_{BW} / \cos \psi} {k_m \sqrt{2SNR}}
\end{equation}
where $\theta_{BW}$ is the 3dB beamwidth in the angular coordinate and $k_m$ is the monopulse pattern difference slope. In addition to SNR, angular measurement error increases when targets are located offset from the beam center due to a loss of gain \cite{Charlish2011_Thesis}, which is reflected in the term $\cos \psi$. $\psi$ is the beam scan angle off-broadside. Similarly, a range error can be represented as zero-mean Gaussian noise with standard deviation given by 
\begin{equation}
\sigma_R \approx \frac {\Delta R}{\sqrt{2SNR}}
\end{equation}
where $\Delta R$ is radar range resolution. 

To make the problem simple, three dimensional position and heading angle $\mathbf{x}^{(i)_s}=[x_i, y_i, z_i, \theta_i]^T$ of each sensing agent $i\in\mathcal{N}$ are assumed to be known. There is no uncertainty about the state of sensing locations. Thus, the information about the target state can be obtained by considering the probability distribution of the target and the induced distribution of measurement variables\footnote{If the sensor state is represented by a probability distribution over the possible sensing locations, the distribution of the measurement variable can be obtained from the observation model. The distribution of the measurement variable usually spread more widely than when the sensor state has no uncertainty, thus the less information about the target states is contained  in the measurement.}.  % 센서의 상태에 불확실성이 들어간다면 얻는 information이 적어지는 효과가 있을 것이다. 전체적인 문제의 formulation에는 큰 변화가 없을 것이다는 요지로 내용 추가. 

\subsection{Estimation}
Estimation of multi-target tracking is much harder than single target tracking, because the number of targets varies over time and the measurements need to be assigned to tracks. These issues have their own importance in multi-target tracking problems. However, in this paper, we aim at proposing a planning algorithm to maximize information about the states of moving targets. To focus on the algorithm that decides the next sensing positions we assume that data association is known perfectly and the number of targets to track is known and fixed. With the assumption that each target moves independently from the other targets, the single filter for each target is sufficient to develop the planning method. 

%The two most basic filters for target tracking are the Kalman filter and particle filter \cite{Song2011_Col}. Although the particle filter is well suited to the  nonlinear and/or non-Gaussian models, the Extended Kalman Filter (EKF) . 
The planning algorithm we propose can be applied with various estimation filters. Here, for each target $j$, the tracking process is performed by an Extended Kalman Filter (EKF). In case of well defined transition models, the EKF has been considered to be a practical means of nonlinear state estimation \cite{BarShalom1990_Book}. Two types of targets are simulated for the simulation in this chapter:  a nearly constant velocity model and a Dubins vehicle. When the sampling time is sufficiently small, Dubins car models fit into the target dynamics in (\ref{eq:model_targetdyn_cv}) \cite{Ding2012_ACC}. 

The EKF accomplishes the sequential estimation of the mean and covariance through two stages: prediction and update as follows \cite{Charlish2011_Thesis}
\vspace{.05in}
\begin{itemize}
	\item [--] Prediction\\
	State estimate: $\mathbf{x}_{k|k-1}=F_{k-1}\mathbf{x}_{k-1|k-1}$ \\
	State covariance: $P_{k|k-1} = F_{k-1}P_{k-1|k-1}F'_{k-1} + Q_{k-1}$ \\ 
	Measurement: $\bar{\mathbf{z}}_{k|k-1} = h(\mathbf{x}_{k|k-1},~ \mathbf{x}^s_{k})$\\
	Innovation covariance: $S_k = H_kP_{k|k-1}H'_k + R_k$
	\item [--] Update \\
	Innovation: $\tilde{\mathbf{z}}_k=\mathbf{z}_k- \bar{\mathbf{z}}_{k|k-1}$ \\
	Filter gain: $W_k = P_{k|k-1}H_k'S_k^{-1}$ \\
	State estimate: $\mathbf{x}_{k|k} = \mathbf{x}_{k|k-1}+W_k\tilde{\mathbf{z}}_k$ \\
	State covariance: $P_{k|k} = P_{k|k-1}-W_kS_kW_k'$
\end{itemize} \vspace{.05in}
where $P_{k|k-1}$ and $P_{k|k}$ are the prior and posterior error covariance, respectively. $H_k=\frac{\partial \mathbf{z}_k} {\partial \mathbf{x}_{k|k-1}} $ is the linearized measurement matrix evaluated at the predicted state $\mathbf{x}_{k|k-1}$. Since the state variables are assumed to follow the normal distribution, the induced variables from the dynamic model and the measurement model also follow normal distributions and the joint probability distribution of all the variables including the target states and measurements are denoted by multivariate Gaussian distributions. Therefore, the covariance update formula is a representation of conditional covariance matrix given the measurement variables. In the next section, we will use the modified equation of covariance matrix considering the time step to calculate the amount of information contained in the measurement variables. 

% In this section, we review the literature on multi-target tracking. We start by describ- ing two of the most basic methods—the Kalman filter [19] and particle filter [15]. They are stochastic methods and solve tracking problems by taking the measurement and model uncertainties into account during object state estimation. They have been extensively used in the vision community for tracking, but these methods are not powerful for tracking multiple objects by themselves, e.g., the Kalman filter and particle filter assume a single measurement at each time instant. In [14], particle fil- ters were used to track multiple objects by incorporating probabilistic MHT [31] for data association. We describe the MHT [31] and JPDAF [4] strategies for tracking multiple targets.

\section{Non-myopic Sensor Network Planning}
The variables of interest in target tracking is the whole states of targets denoted by $X^{\mathcal{T}}=\{\mathbf{x}^{(1)}, \dots, \mathbf{x}^{(M)}\}$. The goal of sensor network planning is to find out the sequences of sensing locations for a sensor network such that information about the states of all targets is maximized. Therefore, the objective function can be the quantity of information about the targets' kinematics at the instant of interest. Sensor planning is called myopic when only the states at the next step are considered. While myopic planning has low computational costs and provides good performance in many examples \cite{Krause2008_JMLR, Hoffmann2010_IEEETAC, Choi2007_ACC}, it performs worse than non-myopic scheduling in some cases. \cite{Liu2003_ICASSP} shows that a simple scenario with sensor holes can cause the performance degradation in myopic planning. To overcome the problem of greedy algorithm, a potential game based method that considers the change over a multi-step lookahead horizon will be provided. 

\subsection{Information-Maximizing Planning Framework}
The goal of sensor planning for a target tracking problem is to select the sequences of sensing decisions for the next time steps that have the most information about the target states. The information can be quantified by mutual information, which is the difference between the entropy of the target state and its conditional entropy conditioned on the measurement variables \cite{Cover1991_Book}. Entropy is the measure of uncertainty in a random entity. More specifically, (differential) entropy of a continuous random variable with probability density function $f_{\mathrm{x}}(x)$ is defined as 
\begin{equation}
\mathcal{H}(\mathbf{x}) \triangleq -\mathbb{E}_{\mathbf{x}} [\log f_{\mathbf{x}}(x)] 
\end{equation}
where $\mathbb{E}_{\mathbf{x}} [\cdot]$ denotes expectations over random entity $\mathbf{x}$. The conditional entropy is defined as 
\begin{equation}
\mathcal{H}(\mathbf{x}|\mathbf{z}) = \mathbb{E}_{\mathbf{z}} [\mathcal{H}(\mathbf{x}|\mathbf{z}=z)] 
\end{equation}
$\mathcal{H}(\mathbf{x}|\mathbf{z})$ is an expected entropy of the conditional distribution taken over all possible values of random variable $\mathbf{z}$. It represents a measure of uncertainty that will remain in $\mathbf{x}$ on the average before the specific value of $\mathbf{z}$ is given. From the definition of the conditional entropy we can compute the expected reduction in uncertainty of the target states before the new measurements are taken. 
\begin{equation}
\mathcal{I}(\mathbf{x};\mathbf{z}) = \mathcal{H}(\mathbf{x}) - \mathcal{H}(\mathbf{x}|\mathbf{z}) 
\end{equation}

As shown in Fig. \ref{fig:sec2_prob_setting}, there are $N$ sensors in a sensor network and $M$ targets to track. We seek to maximize the information about the states of targets at $K$ steps later by selecting the most informative sequences of sensing decisions over the next $K$ time steps. 
The global objective of a sensor network can be represented by 
\begin{equation}
\max_{Z^{\mathcal{N}}_{1:K}} ~\mathcal{I}(X^\mathcal{T}_{K};Z^{\mathcal{N}}_{1:K})
\label{eq:multitarget_global1}
\end{equation}
where $X^{\mathcal{T}}_K=\{\mathbf{x}^{(1)}_K, \dots, \mathbf{x}^{(M)}_K\}$ is a set of random variables representing all of the multi-target states at the last time step of the planning horizon. $Z^{\mathcal{N}}_{1:K}=\{\mathbf{z}^{(1)}_{1:K}, \dots, \mathbf{z}^{(N)}_{1:K}\}$ represents a sequence of predicted  measurements to be taken by a sensor network $\mathcal{N}$ over a $K$-step horizon.
All the variables here follow the probability distribution after updating with the previous measurement history, and the time step $k$ of the subscript represents the relative time from the planning time. Thus, the random variables change with the time at which the planning decisions are made. Note that the variables of interest is the states at the final time of the planning horizon, not the sum of rewards resulting from each decision stage. References \cite{Chhetri2006_JASP} and \cite{Williams2007_IEEETSP} adopted the additive cost considering also the cost at intermediate points in time. Even though they showed the good performance for tracking a single target, the summation could be computationally expensive for multi-target tracking. Thus, we consider only the terminal reward of a sensor sequence as our objective to reduce the number of variables included in computing mutual information. Also, since the final expected target states reflect the intermediate condition of the targets, it is sufficient to consider the mutual information about the final states only. 

\begin{figure}[t]
	\centering
	\includegraphics[width=0.8\columnwidth]{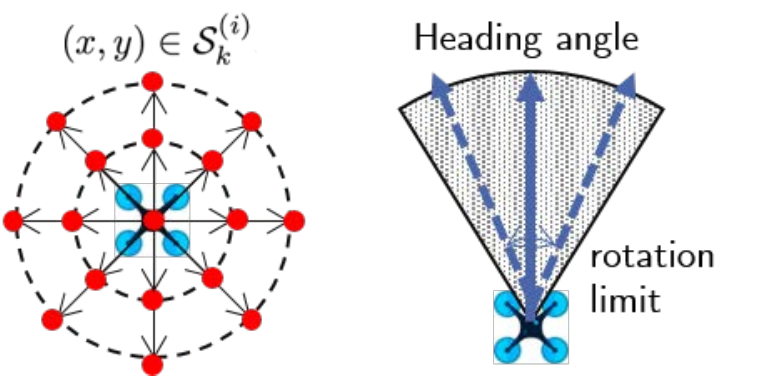}
	\caption{Possible actions for a sensor: locations(left) and heading angle(right), the shaded area in the right figure is the field of view and the two arrows are the limit to rotation of a UAV in one time step.}
	%\vspace{.0in}
	\label{fig:sec3_uavactions}
\end{figure}

The quantity of information that each measurement has about the targets depends on the sensing position and also the heading angle due to the directivity of sensors. Since a sensor can take one measurement at a time, we should decide which target is to be observed. Therefore, there are three decision variables for each measurement $\mathbf{z}_k^{(i)},~ i\in\mathcal{N},~k\in\{1,2,\dots,K\}$, which are represented by a tuple: $a_k^{(i)}\triangleq [x, y, \theta, j]_k^{(i)}$. Here, $(x, y)_k^{(i)}$ denotes the sensing location for sensor $i\in\mathcal{N}$ at time step $k$. As shown in Fig. \ref{fig:sec3_uavactions}, the number of possible sensing locations at each time step is finite and the set of sensing candidates is denoted by $S_k^{(i)}$. For example, in the figure $|S_k^{(i)}|=17$: stay at the previous location or move one of 2 distances in one of 8 directions. In the scenario for simulation, all the sensors are supposed to fly at different heights and maintain their heights to prevent collision between agents, thus $z$-coordinate is not a decision variable. For heading angle of a sensor, there is a limit to the heading angle that can be rotated in one time step, and due to limited field of view of sensors the heading angle is important for success of taking measurements. When it is determined which target is observed, the best heading angle to the target can be decided from the current pose so that it does not exceed the limit of rotation. The heading angle is considered as a function of the current sensor's pose, next sensing candidate location, and selected target's estimated position. Therefore, the pose of a sensor can be dropped out of the decision variables. The optimization problem (\ref{eq:multitarget_global1}) can be rewritten with the modified decision variables as 
\begin{equation}
{a^{\mathcal{N}}_{1:K}}^* = \arg \max_{a^{\mathcal{N}}_{1:K}} ~\mathcal{I}(X^\mathcal{T}_{K};Z^{\mathcal{N}}_{1:K})
\end{equation}
where ${a^{\mathcal{N}}_{1:K}}$ is the set of $a^{(i)}_{k}$ for all $i\in\mathcal{N}$, $k\in\{1,2,\dots, K\}$. Each action for sensor $i$ at time step $k$ is denoted by $a^{(i)}_{k}=[(x,y), j]$ where $j\in\mathcal{T}$ and $(x,y)\in S_k^{(i)}$.

The above objective function for a multi-target tracking problem can be simplified when the targets are assumed to move independently of each other, because the state of each target becomes independent of other targets. 

\begin{lemma}
	\label{lem:multitarget_global}
	If each target moves independent of each other, the objective function (\ref{eq:multitarget_global1}) can be rewritten as a sum of information between each target and the measurements that are taken from that target, 
	\begin{equation}
	\mathcal{I}(X^\mathcal{T}_{K};Z^{\mathcal{N}}_{1:K}) = \sum_{j\in\mathcal{T}} \mathcal{I}(\mathbf{x}_K^{(j)}; Z_1^{N_{j,1}}, \dots, Z_K^{N_{j,K}}) 
	\end{equation}
	where $N_{j,k}\subset\mathcal{N}$ is a subset of sensors which decide to observe target $j$ at time step $k$.  $Z_1^{N_{j,1}}, \dots, Z_K^{N_{j,K}}$ are the set of measurement variables that are taken for target $j$ over the planning horizon of length $K$.  
	\begin{proof}
		The above results can be easily induced by applying chain rule of mutual information \cite{Cover1991_Book}. 
		\begin{eqnarray*}
			\mathcal{I}(X^\mathcal{T}_{K};Z^{\mathcal{N}}_{1:K}) &=& \sum_{j\in\mathcal{T}} \mathcal{I}(\mathbf{x}^{(j)}_{K};Z^{\mathcal{N}}_{1:K}| \mathbf{x}^{(1)}_{K}, \dots, \mathbf{x}^{(j-1)}_{K}) \\
			& = & \sum_{j\in\mathcal{T}} \mathcal{I}(\mathbf{x}^{(j)}_{K};Z^{\mathcal{N}}_{1:K}) 
		\end{eqnarray*}
		Applying the chain rule to the measurement variables, then mutual information for each target becomes 
		\begin{equation*}
		\mathcal{I}(\mathbf{x}^{(j)}_{K};Z^{\mathcal{N}}_{1:K}) = \mathcal{I}(\mathbf{x}^{(j)}_{K};Z^{N_{j,1}}_1, \dots, Z^{N_{j,K}}_K) 
		\end{equation*}
		This follows directly from that mutual information between independent variables are zero and the measurement variables from other targets are also independent.
	\end{proof}
\end{lemma}

From the above lemma, the global objective can be obtained by considering information terms separately for each target.

% TODO:  Add resource constraint optimization framework 
\subsection{Potential Game Formulation}
In \cite{Choi2015_IEEETCST}, we presented a potential game approach for selecting the informative sensing points over the next time step and showed that the proposed method can achieve close-to-optimal solution quality by exploiting a systematic decision update procedure of a potential game. In this section, we will extend the game-theoretic framework for cooperative selection to non-myopic sensor planning that decides sensing points over several time steps. Before stating the potential-game formulation for non-myopic planning, a brief introduction to the basics of a potential game will be given first. 

A general strategic form game consists of a finite set of players $\mathcal{P}=\{1,2,\dots,L\}$; each player $i\in\mathcal{P}$ has a finite action set $\mathcal{A}_i$ that the player can select, and has a preference structure over the actions according to its utility (payoff) function $u_i: \mathcal{A} \rightarrow \mathbb{R}$. $\mathcal{A}=\prod_{i\in\mathcal{P}} \mathcal{A}_i$ is a set of all possible combinations of actions for all players to choose at a time. $a=(a_1,a_2,\dots,a_L)\in\mathcal{A}$ is the collection of strategies of all players, called a strategy profile, where $a_i\in\mathcal{A}_i$ denotes the strategy chosen by player $i\in\mathcal{P}$. For notational convenience, $a_{-i}=(a_1,\dots,a_{i-1},a_{i+1},\dots,a_L)$ denotes the collection of actions of players other than player $i$. With this notation, a strategy profile is expressed as $a=(a_i,a_{-i})$.  

A potential game is a non-cooperative game in which the incentive of the players changing their actions can be expressed by a single function, called the potential function. That is, that the player tries to maximize its utility is equivalent to maximizing the global objective global objective \cite{Monderer1996_Games}.
\begin{definition}[Potential Games]
	\label{def:PG}
	A finite non-cooperative game $\mathcal{G}=\langle \mathcal{P}, \{\mathcal{A}_i\}_{i\in\mathcal{P}}, \{u_i\}_{i\in\mathcal{P}}  \rangle$ is a potential game if there exists a scalar function $\phi:\mathcal{A}\rightarrow\mathbb{R}$ such that \\
	\begin{equation}
	\label{eq:def:PG}
	u_i(a_i',a_{-i})-u_i(a_i'',a_{-i})=\phi(a_i',a_{-i})-\phi(a_i'',a_{-i})
	\end{equation}
	for every $i\in\mathcal{P}$, $a_i',a_i''\in\mathcal{A}_i$, $a_{-i}\in\mathcal{A}_{-i}$. The function $\phi$ is referred to as a potential function of the game $\mathcal{G}$.
\end{definition}
The property of a potential game in (\ref{eq:def:PG}) is called perfect alignment between the potential function and the player's local utility functions. In other words, if a player changes its action unilaterally, the amount of the change in its utility is equal to the change in the potential function.

Potential games have two important properties due to the utility alignment \cite{Candogan2011_MathOR}. First one is that the existence of pure strategy Nash equilibrium is guaranteed. Since in a potential game the joint strategy space is finite, there always exists at least one maximum value of the potential function. This strategy profile maximizing the potential function locally or globally is a pure Nash equilibrium. 
Hence, every potential game possesses at least one pure Nash equilibrium. 
\begin{definition}[Nash Equilibrium]
	A strategy profile $a^*\in\mathcal{A}$ is called a (pure) Nash equilibrium if 
	\begin{equation}
	\label{eq:def:NE}
	u_i(a_i^*,a_{-i}^*)\geq u_i(a_i,a_{-i}^*)
	\end{equation}
	for every $a_i\in\mathcal{A}_i$ and every player $i\in\mathcal{P}$.
\end{definition}

The second important property is about the dynamics of a game. Learning algorithm is a process finding out a Nash equilibrium by repeating a game. Many learning algorithms for potential games are established and proven to have guaranteed asymptotic convergence to a Nash equilibrium \cite{Marden2009_IEEETSMC}. 

A potential game approach for sensor planning problems is to design the components of a game while satisfying perfect alignment\cite{Marden2009_IEEETSMC}. In \cite{Choi2015_IEEETCST}, we set each sensing agent as a player participating in a game and adopted a conditional mutual information as a local utility function. When the same player concept is applied to non-myopic planning game, it causes some computational burden. In case of $L$ sensing allowable options for each step, there are a total of $L^K$ distinct sensing sequences of length $K$, resulting in a total of $L^K \times M^K$ actions for each player. A participant then has an exponentially increased number of action strategies with the length of planning horizon. Thus, it is computationally intractable to calculate the utilities for all the different sensing selections. To address this problem, we set a sensor at each time step as a player, yielding a total of $N\times K$ players each having a strategy space $S_k^{(i)}\times \mathcal{T}$. If all the sensing agents have the same possible moving options, the number of actions for each player to consider is constant with the number of time steps the algorithm looks ahead. Instead of exponential increase in a number of actions for each player, this specification of players linearly increases  the number of participants of a game. In summary, 

\begin{figure*}[t]
	\centering
	\includegraphics[width=1.2\columnwidth]{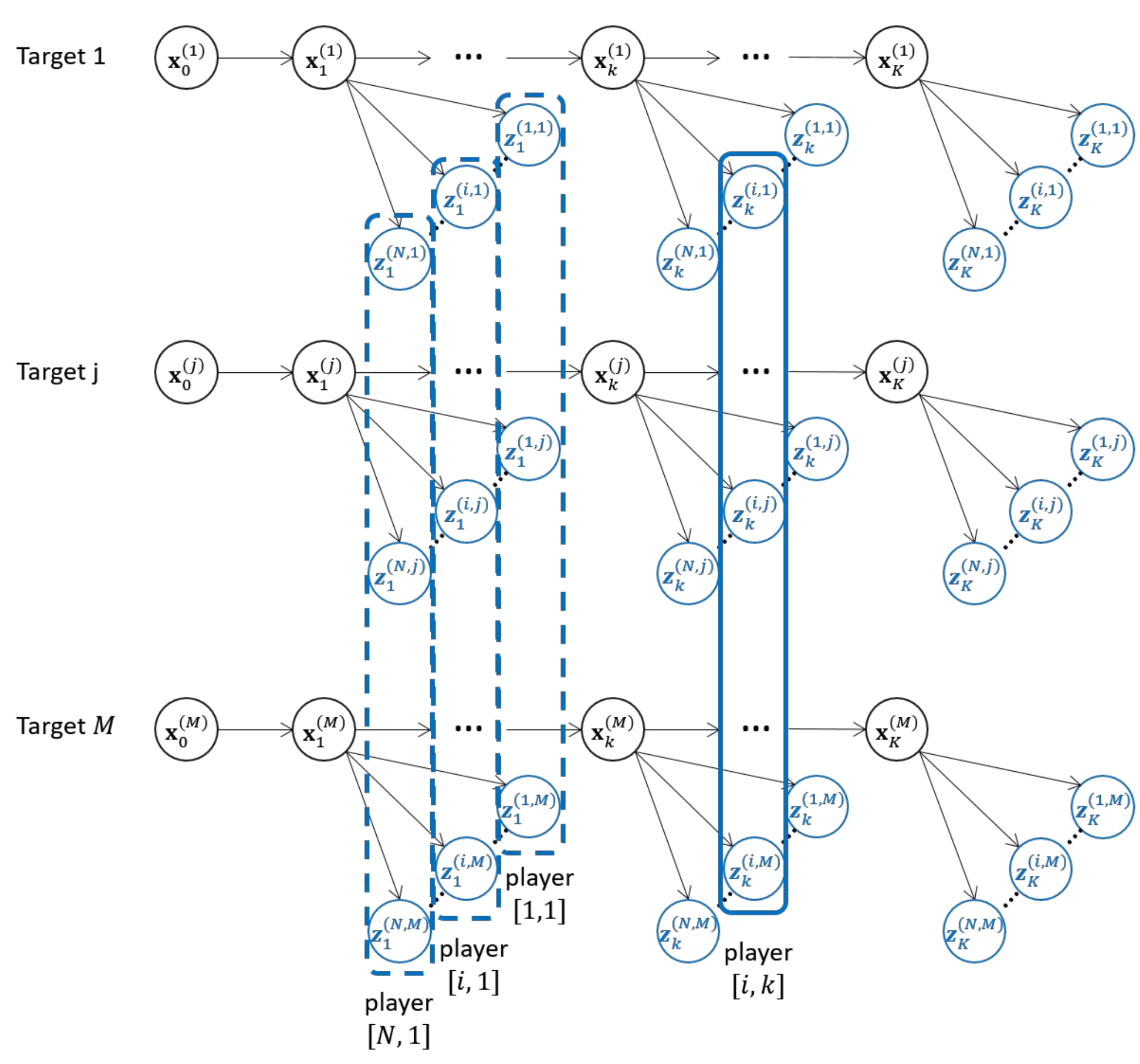}
	\caption{A graphical model with a proposed game design of a multi-target tracking problem in which each target moves independent of each of the targets. A rectangle represent a player in the potential game formulated for cooperative sensor network planning.}
	\label{fig:sec3_graphmodel}
\end{figure*}

%\begin{figure}[t]
%	\centering
%	\includegraphics[width=1.\columnwidth]{sec3_multitarget_graphmodel.pdf}
%	\caption{A graphical model with a proposed game design of a multi-target tracking problem in which each target moves independent of each of the targets. A rectangle represent a player in the potential game formulated for cooperative sensor network planning.}
%	\label{fig:sec3_graphmodel}
%\end{figure}

\begin{itemize}
	
	\item The players are the sensors at each time step represented by the set $\mathcal{P}\triangleq\mathcal{N} \times \mathcal{K}$, where $\mathcal{K}=\{1, \dots, K\}$ is the set of time steps that the planning algorithm looks ahead. Each player is denoted by a 2-tuple $[i, k]$, where $i\in\mathcal{N}, k\in\mathcal{K}$
	\item The strategy space of each player is represented by the set $\mathcal{A}_k^{(i)}\triangleq S_k^{(i)}\times \mathcal{T}$ for each player $(i,k)\in\mathcal{P}$. The action which player $[i,k]$ can select is denoted by the tuple $a_k^{(i)}\triangleq [(x, y), ~j]$, where $(x, y)\in S_k^{(i)}$ is one of the possible sensing locations to which sensor $i$ can move at time step $k$. The candidate positions are constrained by the sensor's moving capability such as maximum speed. The action set for player $[i,k]$ corresponds to the decision variables for measurement $\mathbf{z}_k^{(i)}$ as explained in the previous section. 
\end{itemize}
A graphical model of the state and measurement variables and the players for a potential game formulation is depicted in Fig. \ref{fig:sec3_graphmodel}. 

As proposed in \cite{Choi2015_IEEETCST}, the conditional mutual information conditioned on other players' decisions is considered as a local utility for each agent 
\begin{equation}
u_k^{(i)}=\mathcal{I}(X^\mathcal{T}_{K};\mathbf{z}_k^{(i)}|Z_{-[i,k]})
\end{equation}
where $Z_{-[i,k]}$ is the collection of measurements of players other than player $[i,k]$. As independence between the target states and the measurements from other targets simplifies the mutual information defined as a global objective, the local utility can be rewritten in a simpler form. 

\begin{equation}
u_k^{(i)}=\sum_{j\in\mathcal{T}} \mathcal{I}(\mathbf{x}^{(j)}_{K};\mathbf{z}_k^{(i)}|Z_{-[i,k]})
\label{eq:multitarget_local}
\end{equation}

Since each sensor take measurements of one target at a time, the above expression for a local utility becomes 
\begin{equation}
u_k^{(i,j)}(a_k^{(i)},~a_{-[i,k]})= \mathcal{I}(\mathbf{x}^{(j)}_{K};\mathbf{z}_k^{(i,j)}|Z_{-[i,k]}^{(j)})
\label{eq:multitarget_local1}
\end{equation}
which represents utility value when player $[i,k]$ selects target $j$ for the next measurements. $Z_{-[i,k]}^{(j)}$ represents other measurement variables which are taken for target $j$. Using zero mutual information between two independent variables, the above expression can be easily derived. 
Therefore, it is sufficient to consider the measurements of sensors which select the same target when each player computes its utility values for a specific target selection over possible sensing locations. 

\begin{lemma}
	\label{lem:multitarget_pg}
	With local utility function defined as (\ref{eq:multitarget_local}) the distributed procedure leads to a potential game with a global objective (potential function) as
	\begin{equation*}
	\phi(a_k^{(i)},~a_{-[i,k]})=\mathcal{I}(X^\mathcal{T}_{K};Z^{\mathcal{N}}_{1:K})
	\end{equation*}
	\begin{proof}
		Follows directly from Lemma \ref{lem:multitarget_global} and the proof of Lemma 1 in \cite{Choi2015_IEEETCST}. 
	\end{proof}
\end{lemma}

\section{Learning Algorithm for Non-myopic Sensor Planning}
With the proposed utility function, the designed potential game is solved by a process called a \textit{repeated game}, in which the same set of games is played over and over again while adapting the decisions of players to updated set of information. At each play in a repeated game, every agent chooses an action according to a specific rule (such as fictitious play, better/best response and log-linear learning \cite{Fudenberg1998_Book}), which is generally represented as a probability distribution over the player's actions set. In \cite{Choi2015_IEEETCST}, we adopted the joint strategy fictitious play (JSFP) for finding out Nash equilibrium in a myopic sensor network planning problem formulated as a potential game. In this section, we will present learning dynamics that exploits the same learning framework as in our previous work.  First, we review the basic concept of JSFP summarizing brief description of \cite{Choi2015_IEEETCST}. Next, we introduce the extended learning algorithm to non-myopic planning and analyze the computation complexity to perform the procedure. 

\subsection{Joint Strategy Fictitious Play}
In the fictitious play based algorithms, each players presumes that the opponents choose the action according to the empirical frequency of play. Especially in JSFP, a player keeps track of the joint actions of all others and computes an expected utility based on the joint empirical frequency \cite{Marden2009_IEEETAC}. Let $f_{-i}(a_{-i};t)$ be the frequency with which all players but $i$ have selected joint action profile $a_{-i}$ up to the $t-$th play. Then the expected utility for action $a_i\in\mathcal{A}_i$  is given by
\begin{eqnarray*}
	u_i(a_i;t)&=&\mathbb{E}_{f_{-i}(t)}[u_i(a_i,a_{-i})] \\
	&=& \frac{1}{t} \sum_{\tau=1}^{t-1} u_i(a_i,a_{-i}(\tau))
\end{eqnarray*}
where $u_i(a_i,a_{-i}(\tau))$ is the utility computed for $a_i\in\mathcal{A}_i$ and the joint actions of others $a_{-i}(\tau)$ selected at the previous play $\tau$. In \cite{Marden2009_IEEETAC}, Marden et al gave the simple expression for  the expectation $u_i(a_i; t) $ as 
\begin{equation*}
u_i(a_i;t)=\frac{t-1}{t}u_i(a_i;t-1) + \frac{1}{t}u_i(a_i,a_{-i}(t-1))
\end{equation*}
This recursion form reduces computation complexity significantly compared to the traditional fictitious play based on empirical frequencies of marginal actions \cite{Choi2015_IEEETCST}. Also, with some notion of inertia, i.e., a probabilistic reluctance to change actions, JSFP was shown to converge to a pure strategy Nash equilibrium [\cite{Marden2009_IEEETAC}, Theorem 2.1].

\subsection{Learning Algorithm with JSFP}

\begin{algorithm*} 
	\caption{{\sc JSFP-based Double Loop Learning for Non-myopic Sensor Network Planning}  \\
		INPUT: target states: $\{\mathbf{x}_0^{(1)}, \dots, \mathbf{x}_0^{(M)} \}$, state covariances: $\{P_0^{(1)}, \dots, P_0^{(M)}\}$,
		sensors' pose: $\{ \mathbf{x}_0^{(1)_s}, \dots, \mathbf{x}_0^{(N)_s}\}$, $\alpha \in (0,1)$}
	\begin{algorithmic}[1]
		\STATE Predict target states and covariances $\{\mathbf{x}_{1:K|0}^{(1)}, \dots, \mathbf{x}_{1:K|0}^{(M)} \}$, $\{P_{1:K|0}^{(1)}, \dots, P_{1:K|0}^{(M)}\}$
		\STATE Initialize $N_{j,k}=\emptyset$, $\forall j\in\mathcal{T}$, for $k\in\mathcal{K}$ 
		\STATE ${\tt Convergedout} = {\tt FALSE}$
		\WHILE { $\neg~ {\tt Convergedout}$}
		\FOR{ $k \in \{K,\dots, 1\}$}
		\STATE $S_k^{(i)} =$ generate sensing locations at time step $k$, for each $ i\in\mathcal{N}$ 
		\STATE Update covariance matrix  $\{P_{K|N_{1,-k}}^{(1)}, \dots, P_{K|N_{M,-k}}^{(M)}\}$
		\STATE Initialize $\bar{u}_k^{(i,j)}$ to zero vector 
		\STATE ${\tt Exitin} = {\tt FALSE}$
		\STATE $t:=0$ \\
		\WHILE { $\neg~ {\tt Exitin} $}
		
		\IF {${\tt rand()} > \beta$}
		\STATE ${\tt Exitin} = {\tt TRUE}$
		\ENDIF
		
		\STATE $t:=t+1$
		\FOR {$i\in\mathcal{N}$}
		\STATE $u_k^{(i,j)}=$ compute utility values at time step $k$, for each $(x,y)\in S_k^{(i)}$ and $j\in\mathcal{T}$ 
		\STATE $\bar{u}_k^{(i,j)} :=  \frac{t-1}{t} \bar{u}_k^{(i,j)}+\frac{1}{t}u_k^{(i,j)}$
		\ENDFOR
		\FOR {$i\in\mathcal{N}$}
		\STATE $[(x,y), j]_{(i,k)}^*=\arg \max_{(x,y)\in S_k^{(i)}, j\in\mathcal{T}} \bar{u}_k^{(i,j)} $
		\IF {${\tt rand()} > \alpha \land \neg~ {\tt Exitin}$ }
		\STATE $a_k^{(i)} = [(x,y), j]_{(i,k)}^*$ update strategy to the optimal decision. 
		\ENDIF
		\ENDFOR
		\STATE {${\tt Exitin =}$ check convergence of inner game}
		\ENDWHILE
		\STATE Update $N_{j,k}$ from $a_k^{(1)}, \dots, $ $a_k^{(N)}$ 
		\ENDFOR %{ $k \in \{K,\dots, 1\}$}
		\STATE {${\tt Convergedout =}$ check convergence of outer game}
		\ENDWHILE % Convergedout	
	\end{algorithmic}
	\label{alg:jsfp_nonmyopic}
\end{algorithm*}

In \cite{Choi2015_IEEETCST}, we applied JSFP algorithm to obtain a solution of a myopic sensor planning problem. 
The previously proposed JSFP method was simple in that the method does not need to consider a kinematic constraint of a mobile sensor over multiple time steps. It is sufficient to consider the set of next possible sensing locations only, which is unchanged during a repeated game. However, when considering the path over more than one time step, the $k$-th possible sensing locations and heading angles are subject to the previous and next pose of the sensing platform. Therefore, we should reflect changing reachable candidates at each time step in a learning algorithm. 

Another difficulty in non-myopic planning is sensor holes. For example, when a sensor has a limited field of view and all targets locate out of view at the planning time and in one time step later, a greedy method can get trapped at local maximum \cite{Liu2003_ICASSP}. Extending the candidate region to the $K$-step reachable region, this sensor hole problem can be addressed. Since a mobile platform can reach any locations inside the farthest points reachable by $K$ movements, it is enough to define the set of possible sensor configuration options for the $K$th step candidates as the region including the farthest points and inside that boundary. 

A non-myopic learning algorithm to be presented starts a repeating game from the last time step $K$. Player $[i, K]$ for each $i\in\mathcal{N}$ first selects the optimal position as an initial action within $K$-step reachable region by solving local greedy method. This initialization scheme continues to $(K-1)$th step with conditioning pdf of variables conditioned on the $K$th decisions. After converging to the solution for each time step, the algorithm goes back to the previous time step until the one step lookahead decision. This initialization planning method is similar to dynamic programming. 

In Algorithm \ref{alg:jsfp_nonmyopic}, the learning algorithm for non-myopic planning is summarized. There are two {\tt while} loops: outer and inner. The inner loop corresponds JSFP for a potential game considering $k$-th step only. The inner loop terminates according to the following rules: 
\begin{itemize}
	\item If JSFP converges to a Nash equilibrium, the decisions for $k$-th step are made with those optimal sensing locations.
	\item Otherwise, the loop terminates with probability $1-\beta$, $\beta \in (0,1)$.
\end{itemize}
where $\beta$ represents the willingness to optimize at time step $k$. According to these rules, players try to optimize their decisions with probability $\beta$ and will stay with the previous action with probability $1-\beta$ at each repeated game. This nonzero inertia term is required to avoid cycling in local search \cite{Marden2009_IEEETAC}.
After the inner loop terminates, the selected decisions are stored in $N_{j,k}$'s which are used for updating the covariance matrix. The outer loop is performed recursively until all of the decisions are not changed over a repeated game. Since the algorithm is a variant of JSFP, it can be proved that the algorithm with inertia almost surely converges to a pure strategy Nash equilibrium.  

\begin{theorem}
	\label{thm:multitarget_learning}
	Algorithm 1 with $\alpha \in (0, 1)$ and $\beta \in (0,1)$ almost surely converges to a pure strategy Nash equilibrium of the potential game in Lemma \ref{lem:multitarget_pg}, with consistent tie-breaking in all the arg max operations involved in the process.
\end{theorem}
We provide a proof of Theorem \ref{thm:multitarget_learning} in Appendix. The structure of Algorithm \ref{alg:jsfp_nonmyopic} is similar to fading memory JSFP with inertia, because the utility values are not stored through an outer loop. The utility $\bar{u}_k^{(i,j)}$ is initialized to zero vector every time an inner loop ends. Therefore, we will prove Theorem \ref{thm:multitarget_learning} by following a structure to the proof of Theorem 3.1 in \cite{Marden2009_IEEETAC}. %We encourage the reader to first review the proof of fading memory JSFP with inertia in Theorem 3.1 of the following sec- tion.

%TODO: Add theorem for the convergence of the proposed JSFP-based learning algorithm. 
%		  : Modify the convergence condition to escape the inner loop. 

%______________________________________
\subsection{Calculation of Utility Functions}
This section explains how to compute the utility values in Algorithm \ref{alg:jsfp_nonmyopic} and analyzes its computation cost. 
In the previous section, the local utility function for a multi-target tracking problem is defined as (\ref{eq:multitarget_local}). When the dynamics and measurement models are linear and Gaussian, the mutual information objective of a normal distribution relates to its covariance matrix only. 
\begin{align}
\mathcal{I}(\mathbf{x}^{(j)}_{K}&;\mathbf{z}_k^{(i,j)}|Z_{-[i,k]}^{(j)}) \notag\\ 
&= \mathcal{H}(\mathbf{z}_k^{(i,j)}|Z_{-[i,k]}^{(j)}) - \mathcal{H}(\mathbf{z}_k^{(i,j)}|\mathbf{x}^{(j)}_{K},~Z_{-[i,k]}^{(j)}) \notag \\ 
&= \frac{1}{2}\log \left(\left|P(\mathbf{z}_k^{(i,j)}|Z_{-[i,k]}^{(j)})\right|\right) \notag \\
&~~~~~~~- \frac{1}{2}\log \left(\left|P(\mathbf{z}_k^{(i,j)}|\mathbf{x}^{(j)}_{K},~Z_{-[i,k]}^{(j)})\right|\right)
\label{eq:multitarget_calutil1}
\end{align}
where $P(\mathbf{z}_k^{(i,j)}|Z_{-[i,k]}^{(j)})$ and $P(\mathbf{z}_k^{(i,j)}|\mathbf{x}^{(j)}_{K},~Z_{-[i,k]}^{(j)})$ represent the conditional covariance matrices of $\mathbf{z}_k^{(i,j)}$ conditioned on $Z_{-[i,k]}^{(j)}$ and additional $\mathbf{x}^{(j)}_{K}$, respectively.  When $P(\mathbf{x},\mathbf{z})$ represents the covariance between $\mathbf{x}$ and $\mathbf{z}$, the conditional covariance matrix for a Gaussian can be computed as  
\begin{equation} 
P(\mathbf{x}|\mathbf{z})=P(\mathbf{x})-P(\mathbf{x},\mathbf{z})P^{-1}(\mathbf{x})P(\mathbf{z},\mathbf{x}).
\label{eq:condcov_gauss}
\end{equation}
From the above equation, we need the covariance matrix relating three variables $\mathbf{x}^{(j)}_{K}$, $\mathbf{z}_k^{(i,j)}$ and $Z_{-[i,k]}^{(j)}$ to compute the utility (\ref{eq:multitarget_calutil1}). Here, $\mathbf{x}^{(j)}_{K}$ represents the \textit{a priori} state estimate predicted at the planning time, then 
\begin{equation*}
\mathbf{x}^{(j)}_{K}=\mathbf{x}^{(j)}_{K|0}=F^{(j)}_{K}\mathbf{x}^{(j)}_{0}
\end{equation*}
For notational simplicity, we will omit the superscript $(j)$ specifying the target in this section, then all the variables here are for the single target $j$. Likewise, the measurement variables to be taken at time step $k$ are induced from the a priori estimate $\mathbf{x}^{(j)}_{k|0}$. From target dynamics (\ref{eq:model_targetdyn}) and measurement model
\begin{eqnarray}
\mathbf{x}_{k}=\mathbf{x}_{k|0}=F_{k}\mathbf{x}_{0} + w_k \notag \\ 
\mathbf{z}^{(i)}_{k}=h(\mathbf{x}_{k|0}, \mathbf{x}_k^{(i)}) + v_k^{(i)} \notag
\end{eqnarray}
The measurement noise $v_k^{(i)}$ is a zero mean uncorrelated Gaussian process with covariance $R_k^{(i)}$ and is uncorrelated with any other sensor noise as well as the state variables and the process noise. The covariance matrix are computed as follows, 
\begin{align*}
&P(\mathbf{x}_{k}, \mathbf{x}_{k}) &= &F_k P_{0} F_k' + Q_k ~~~ &\text{for}~ k\in\mathcal{K}\\
&P(\mathbf{z}^{(i)}_{k}, \mathbf{z}^{(i)}_{k}) &= &H_k^{(i)} P(\mathbf{x}_{k}, \mathbf{x}_{k}) {H^{(i)}_k}'  + R_k^{(i)} ~~&\text{for}~ k\in\mathcal{K}\\
&P(\mathbf{z}^{(i)}_{k}, \mathbf{x}_{K}) &= &H_k^{(i)} P(\mathbf{x}_{k}, \mathbf{x}_{k}) F_{K-k} '&\\ 
&P(\mathbf{z}^{(i)}_{k}, \mathbf{z}^{(i)}_{l}) &=& H_k^{(i)} P(\mathbf{x}_{k}, \mathbf{x}_{k}) F_{l-k}' {H_l^{(i)}}' ~~~&\text{for}~ l<k
\end{align*}
where $F_{l-k}$ is system transition matrix from time step $k$ to $l$ as defined in (\ref{eq:model_targetdyn_cv}) with appropriate time interval between two time steps. The elements in the covariance matrix related to the target states $P(\mathbf{x}_{k}, \mathbf{x}_{k})$ can be obtained prior to executing the learning algorithm for solving the potential game. Predicting the covariance matrices of the target states from $k=1$ to $k=K$ requires $\mathcal{O}(K)$. There are two types of the measurement variables in (\ref{eq:multitarget_calutil1}). While $\mathbf{z}_k^{(i,j)}$ is the measurements for a sensing candidate at the current game, $Z_{-[i,k]}^{(j)}$ are the measurement variables selected at the previous game. Before starting each game in a repeated game, the terms related to the measurement variables can be computed and the resulting conditional covariance matrix $P(\mathbf{z}_k^{(i,j)}|Z_{-[i,k]}^{(j)})$ and $P(\mathbf{z}_k^{(i,j)}|\mathbf{x}^{(j)}_{K},~Z_{-[i,k]}^{(j)})$ are obtained.  The maximum size of $Z_{-[i,k]}^{(j)}$ is $(N-1)(K-1)$. Since inversion of a $n \times n$ symmetric positive matrix requires $\mathcal{O}(n^3)$ flops and a determinant computation using Cholesky factorization needs $\mathcal{O}(n^3)$ flops \cite{Andersen2002_LectureNotes}, the resulting computation time for obtaining a set of utility values of one agent is $\mathcal{O}(N^3 K^3)+\mathcal{O}(L)$.  After obtaining the conditional matrix, computing each utility value requires constant time, resulting in $\mathcal{O}(L)$ because the size of $\mathbf{z}_k^{(i,j)}$ is constant (2 for this scenario). In a repeated game, the prediction of the target states is done once, on the other hand, the conditional covariance matrix is computed every game step. Therefore, the overall computing time is approximated as $\mathcal{O}(N^3 K^3)$ if $L$ is fixed. Compared with myopic planning ($K=1$), the calculation time increases polynomially to the length of the time horizon. 
%As shown in (\ref{eq:model_targetdyn_cv}), both matrices $F_k$ and $Q_k$ depend only on the sampling interval $\Delta t$, thus we can calculate the matrices before running the planning algorithm. 

%Ref
%  \cite{Bertsekas2000_Book, Skoglar2007_Book} a nice property of OLFC is that it performs at least as well as an optimal open-loop policy. 

\section{Numerical Simulations}
\begin{figure}[t]
	\centering
	\includegraphics[width=.7\columnwidth]{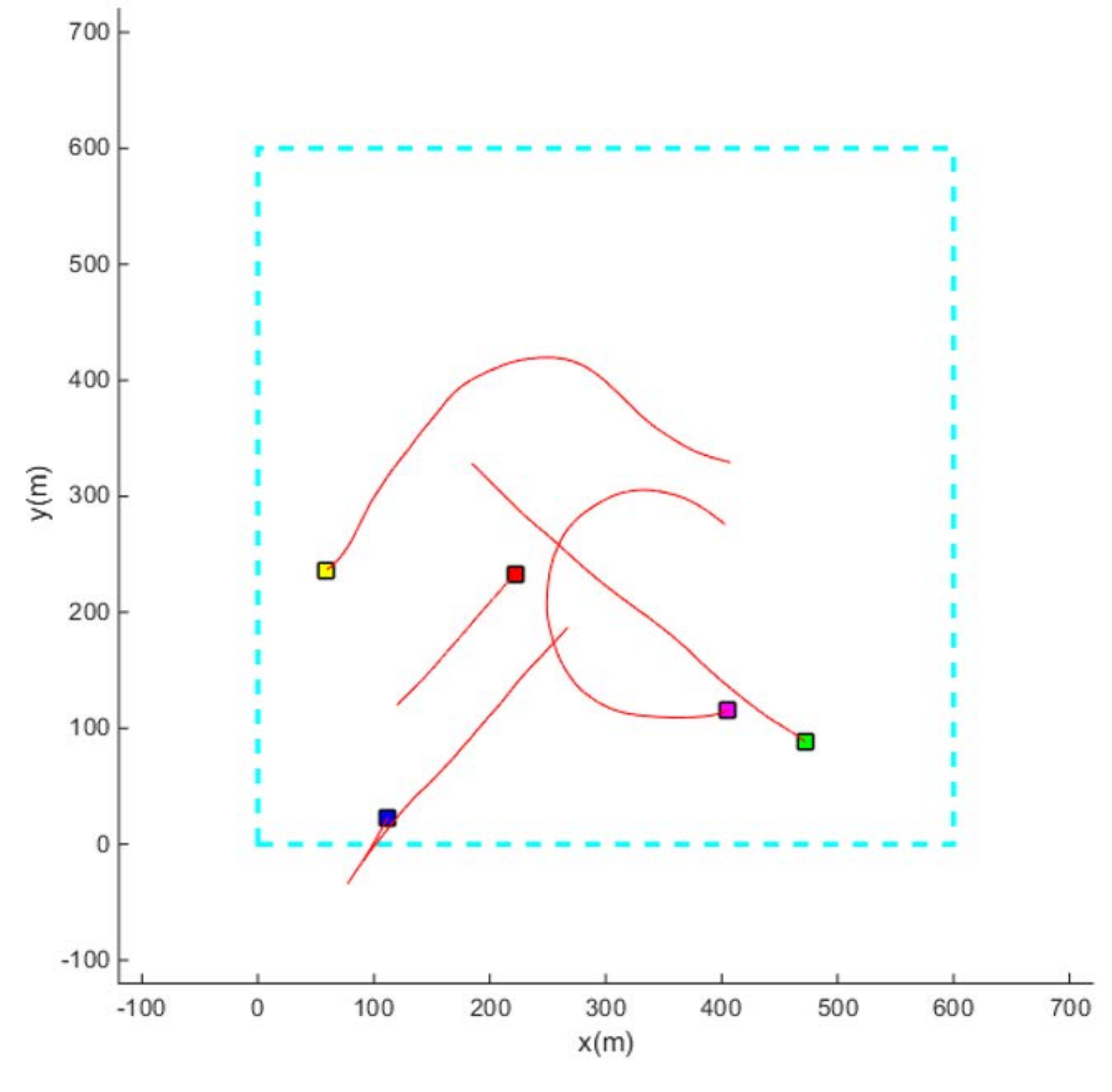}
	\caption{One example of multi-target tracking scenario, in which (red, green, and blue) squares move in nearly constant velocity and the other two squares follow Dubins vehicle model. }
	\label{fig:multitarget_scenario}
\end{figure}

To compare the performance of a new algorithm with a greedy method, a scenario with sensor holes is presented. There are five targets moving on the ground. The map size is approximately $600 \times 600$. Three of targets go straight with nearly constant velocity and change their speed sign when the target crosses the map boundary. The remaining two targets are modeled as Dubins vehicles which have a fixed forward velocity with a bounded turning radius. One example of this scenario is shown in Fig. \ref{fig:multitarget_scenario} (red, green, blue: nearly constant velocity model. magenta, yellow: Dubins vehicles). The reason there are two types of target movements is to see the performance difference between a well predicted model and a poorly matched model. A multi-step lookahead sensor planning is expected to outperform short-term methods in situations where the dynamics of targets are predictably changing, but to give poor performance in cases where the dynamic model is not matched to a real target dynamics. A target model applied to EKF is close to a nearly constant velocity model, while Dubins vehicles are not fit into that dynamics. 

\begin{figure}[]
	\centering
	\includegraphics[width=1\columnwidth]{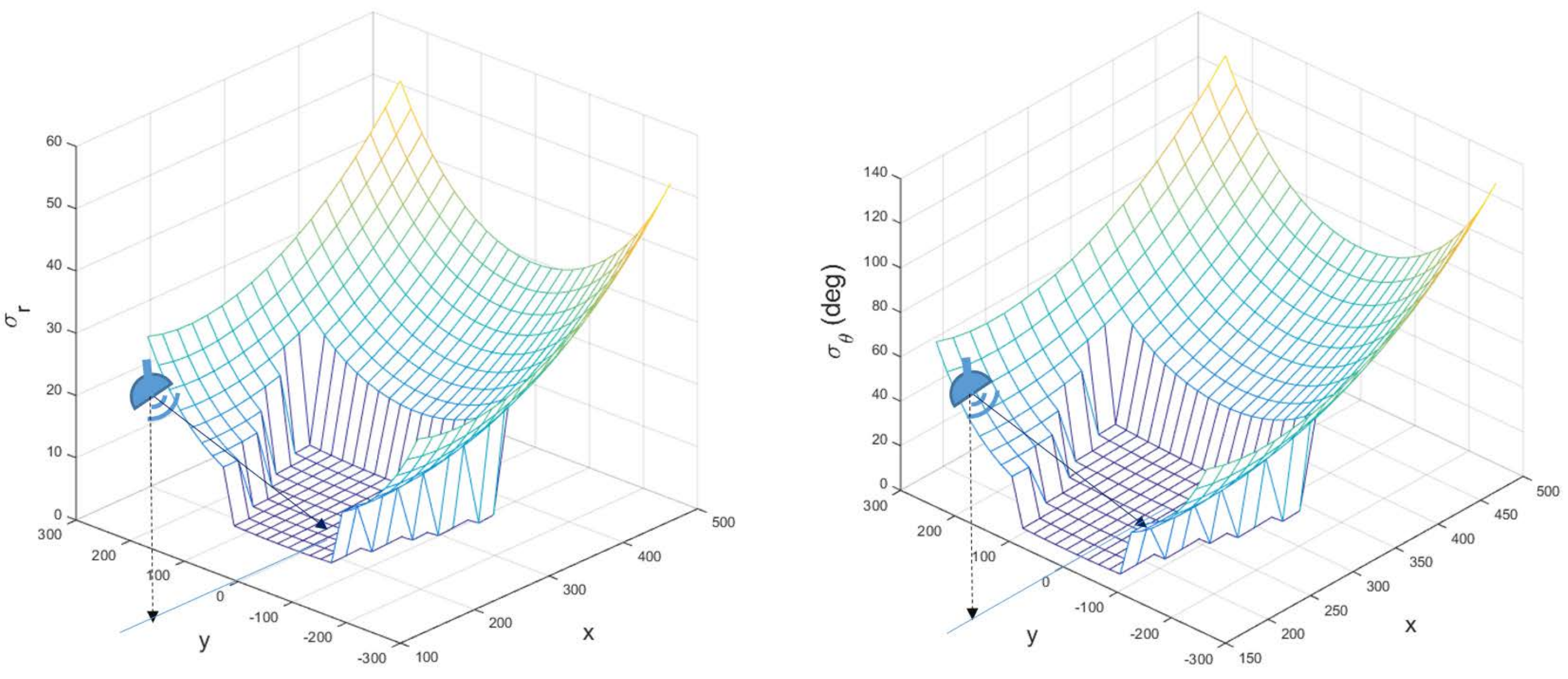}
	\caption{Measurement error term of a radar-like sensor. A sensor locates at (0,0, 490) and points to the positive x-axis slightly looking down.}
	\label{fig:multitarget_sensormodel}
\end{figure}

\begin{figure}[]
	\vspace{-.0in}
	\centering
	\includegraphics[width=.95\columnwidth]{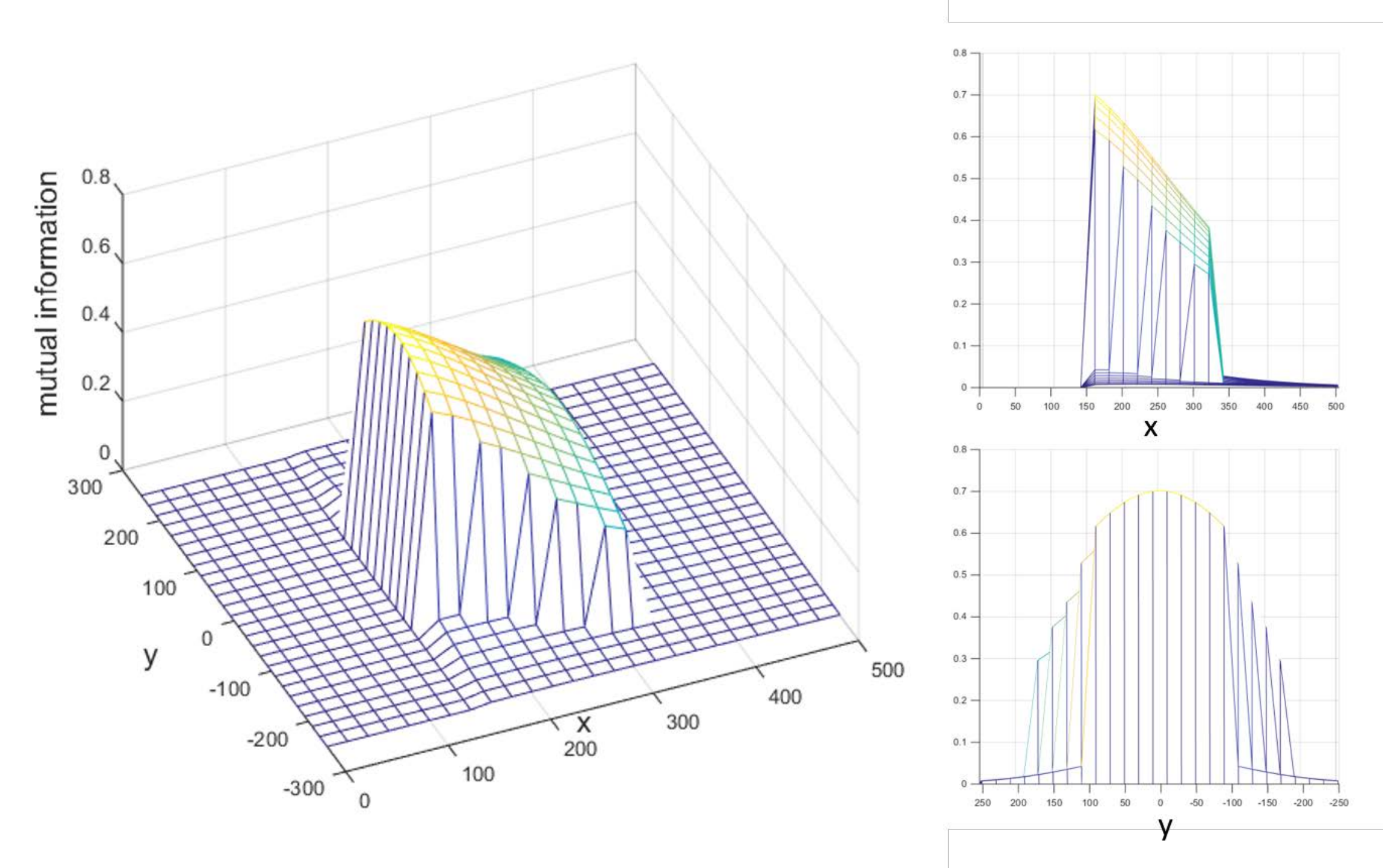}
	\vspace{-.0in}
	\caption{Mutual information with varying target location on the ground.}
	\label{fig:multitarget_mutinf}
\end{figure}

For sensor platforms, we consider a set of quadrotors moving at different heights from each other to alleviate collision avoidance constraints. In a scenario for simulation, there are three platforms, each of which is equipped with one radar-like sensor. The sensors are mounted at a slight angle to look down. The standard deviations of measurement noise are depicted in Fig. \ref{fig:multitarget_sensormodel}, in which a sensor is located 500 high from the ground and $(x,y)=(0,0)$. Fig. \ref{fig:multitarget_mutinf} depicts the mutual information that can be obtained when a ground target is located with respect to the sensor. %, which is considered as the minus log of determinant of the posterior covariance matrix. 
The coordinates of the sensor are $(0,0)$ and the sensor is oriented towards the positive x-axis as shown in Fig. \ref{fig:multitarget_sensormodel}. 

\begin{figure}[]
	\centering
	\includegraphics[width=1\columnwidth]{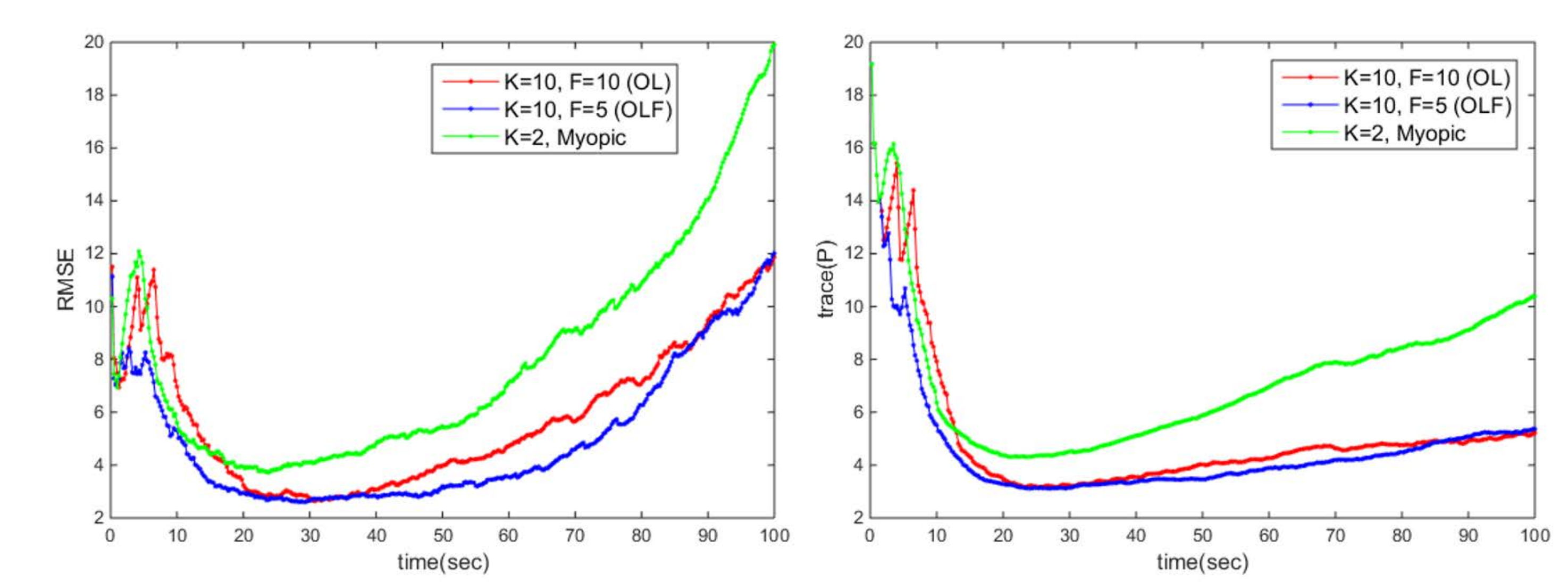}
	\caption{Comparison of the RMSE of the target position for nearly constant velocity target}
	\label{fig:multitarget_resultCV}
	%\vspace{1in} 
	\centering
	\includegraphics[width=1\columnwidth]{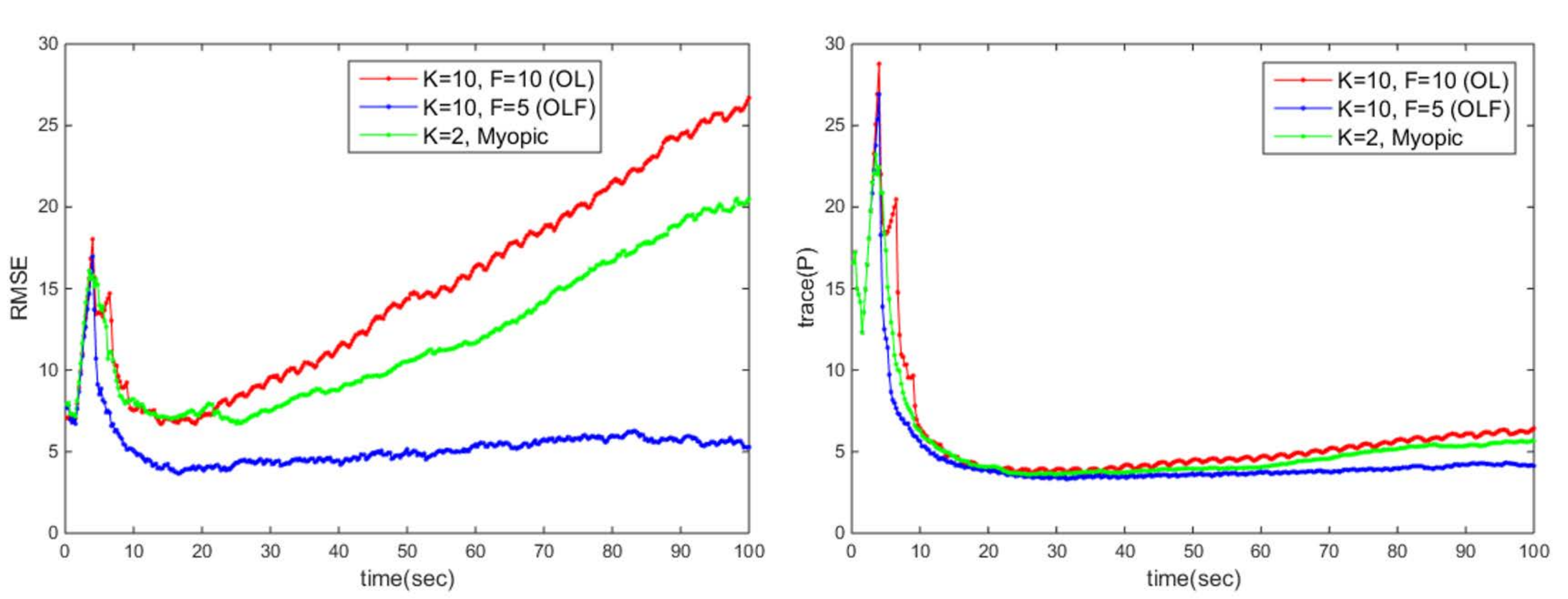}
	\caption{Comparison of the RMSE of the target position for Dubins vehicle target}
	\label{fig:multitarget_resultDubins}
\end{figure}

The proposed non-myopic planning is realized in two ways as \cite{Chhetri2006_JASP}. The first is open-loop planning (OL), in which the planning is performed after all the planned decisions are exhausted. The second is open-loop feedback (OLF), which is same as rolling horizon planning. In OLF, the first some part of decisions is executed, and then a new plan for the following $K$ steps is generated, having updating the covariance matrices and the state variables. To overcome the sensor hole problem and prevent the divergence of the filter, the decisions are made every two time steps and take measurement of the same target as the previous one in the intermediate time step. Thus, $K=2$ corresponds to myopic planning, which finds out the optimal sensing actions in two steps and the same target is observed for two steps. 
We ran 100 simulations using both non-myopic planning schemes. For each run, the states of targets and the positions of the sensors are randomly initialized. Figs. \ref{fig:multitarget_resultCV} and \ref{fig:multitarget_resultDubins} present a comparison between myopic planning and two different non-myopic planning methods. As expected, when the target dynamics is predicted the real target movements well, the performance of non-myopic planning is better than myopic planning (See Fig. \ref{fig:multitarget_resultCV}). On the other hand, when the dynamic model for the target state is not fit into the actual movement of the target, non-myopic planning (OL) gives poorer performance (See Fig. \ref{fig:multitarget_resultDubins}) than a greedy method. However, the open-loop feedback planning overcomes the nonlinearity of the target dynamics and gives the best performance of all the planning methods. 

%-------------------------------------------------------------------%
\section{Conclusions}% and Future Work}
In this paper we have investigated a potential game approach to long-term sensor network planning. Since the non-myopic optimization problem requires exponentially increasing computation time as the length of the planning time horizon increases, a player for a potential game is needed to be defined differently from the myopic case to keep the size of the problem manageable. Accordingly, a new learning algorithm was proposed combining dynamic programming and JSFP.  To demonstrate the performance of the non-myopic planning method, we considered a multi-target tracking problem involving sensor holes due to constrained mobility of platforms and limitations on sensor visibility. The numerical simulations showed the effectiveness of the proposed approach and the conditions in which non-myopic planning gives better performance than myopic planning.

\section*{Appendix. proof of theorem 1} 
According to \cite{Marden2009_IEEETAC}, for the proof of convergence it is sufficient to show that there exists a non-zero probability, $\epsilon^*> 0$, such that %the following statement holds. For any $t \ge 0$, $a(t) \in 
the probability of convergence to an equilibrium by finite time $t^*$ is at least $\epsilon^*$. Since $\epsilon^*$ does not depend on the stage $t$ of a learning, this will imply that the action profile converges to an equilibrium almost surely.

In \cite{Marden2009_IEEETAC}, the proof of fading memory JSFP with inertia relies on the fact that if the current action profile is repeated a sufficient number of times (finite and independent of time) then a best response to the empirical frequencies becomes equivalent to a best response to the current action profile and hence will increase the potential provided that there is only a unique deviator. This will always happen with at least a fixed (time independent) probability because of the inertia. The outer loop of the proposed learning algorithm considers the most recent information only. That is, the outer structure of the learning rule is fading JSFP with inertia considering the very previous decisions only. Thus, an outer loop with the same action profile does not need to be repeated. If there is a probability that one player changes its decision unilaterally to optimize its decision, the proof will be completed. Due to the parameter $\beta$ and the inertia term $\alpha$ of JSFP in an inner loop, this unilateral best response will always happen with at least a fixed probability. Thus, the iterative procedure is proven to converge to a pure strategy Nash equilibrium almost surely as in the proof of the Theorem 3.1 in \cite{Marden2009_IEEETAC}.

	\addtolength{\textheight}{-12cm}   % This command serves to balance the column lengths
	
	%%%%%%%%%%%%%%%%%%%%%%%%%%%%%%%%%%%%%%%%%%%%%%%%%%%%%%%%%%%%%%%%%%%%%%%%%%%%%%%%

	%%%%%%%%%%%%%%%%%%%%%%%%%%%%%%%%%%%%%%%%%%%%%%%%%%%%%%%%%%%%%%%%%%%%%%%%%%%%%%%%

	%%%%%%%%%%%%%%%%%%%%%%%%%%%%%%%%%%%%%%%%%%%%%%%%%%%%%%%%%%%%%%%%%%%%%%%%%%%%%%%%

	\bibliographystyle{IEEEtran}
	\bibliography{manuscript}

\end{document}